\documentclass[twocolumn,amsmath,amssymb,showpacs]{revtex4}
\usepackage{epsfig}
\usepackage{graphicx}

\begin{document}

\title{Isometric group of $(\alpha,\beta)$-type Finsler space and the symmetry of Very Special Relativity }

\author{Xin Li$^{1,4}$}
\email{lixin@itp.ac.cn}
\author{Zhe Chang$^{2,4}$}
\email{zchang@ihep.ac.cn}
\author{Xiaohuan Mo$^{3}$}
\email{moxh@pku.edu.cn}
\affiliation{${}^1$Institute of Theoretical Physics,
Chinese Academy of Sciences, 100190 Beijing, China\\
${}^2$Institute of High Energy Physics, Chinese Academy
of Sciences, 100049 Beijing, China\\
${}^3$Key Laboratory of Pure and Applied Mathematics
School of Mathematical Sciences, Peking University, Beijing 100871, China\\
${}^4$Theoretical Physics Center for Science Facilities, Chinese Academy of Sciences}

\begin{abstract}
The Killing equation for a general Finsler space is set up. It is showed that the Killing equation of $(\alpha,\beta)$ space can be divided into two parts.
One is the same with Killing equation of a Riemannian metric, another equation can be regarded as a constraint. The solutions of Killing equations present explicitly the isometric symmetry of Finsler space.
We find that the isometric group of a special case of $(\alpha,\beta)$ space is the same with the symmetry of Very Special Relativity (VSR). The Killing vectors of Finsler-Funk space are given. Unlike Riemannian constant curvature space, the 4 dimensional Funk space with constant curvature just have 6 independent Killing vectors.
\end{abstract}
\pacs{02.40.-k,11.30.-j}

\maketitle
\section{Introduction}

In the past few years, two interesting theories of investigating the violation of  Lorentz Invariance (LI) are proposed. One is the so called Doubly Special Relativity (DSR) \cite{Amelino1,Amelino2,Amelino3,Smolin1,Smolin2}. This theory takes Planck-scale effects into account by introducing an invariant Planckian parameter in the theory of special relativity. Another is the so called Very Special Relativity (VSR) developed by Cohen and Glashow \cite{Glashow}. This theory suggested that the exact symmetry group of nature may be isomorphic to a subgroup SIM(2) of the Poincare group. And the SIM(2) group semi-direct product with the spacetime translation group gives an 8-dimensional subgroup of the Poincare group called ISIM(2) \cite{Kogut}. Under the symmetry of ISIM(2), the CPT symmetry is preserved and many empirical successes of special relativity are still functioned.

Recently, Physicists found that the two theories mentioned above are related with Finsler geometry. Girelli, Liberati and
Sindoni \cite{Girelli} showed that the Modified dispersion relation (MDR) in DSR can be incorporated into
the framework of Finsler geometry. The symmetry of the MDR was described in the Hamiltonian formalism. Also, Gibbons, Gomis and Pope \cite{Gibbons} showed that the Finslerian line element $ds=(\eta_{\mu\nu} dx^\mu dx^\nu)^{(1-b)/2}(n_\rho dx^\rho)^b$ is invariant under the transformations of the group DISIM$_b(2)$ (1-parameter family of deformations of ISIM(2)).

Finsler geometry as a natural generation of Riemann geometry could provide new sight on modern physics. the model based on Finsler geometry could explain the recent astronomical observations which Einstein's gravity could not. An incomplete list includes: the flat rotation curves of spiral galaxies can be deduced naturally without invoking dark matter \cite{Finsler DM}; the anomalous acceleration\cite{Anderson} in solar system observed by Pioneer 10 and 11 spacecrafts could account for a special Finsler space-Randers space \cite{Finsler PA}; the secular trend in the astronomical unit\cite{Krasinsky,Standish} and the anomalous secular eccentricity variation of the Moon's orbit\cite{Williams} could account for effect of the length change of unit circle in Finsler geometry\cite{Finsler AU}.

Thus, the symmetry of Finslerian spacetime is worth investigating. The way of describing spacetime symmetry in a covariant language (the symmetry should not depend on any particular choice of coordinate system) involves the concept of isometric transformation. In fact, the symmetry of spacetime is described by the so called isometric group. The generators of isometric group is directly connected  with the Killing vectors\cite{Killing}. In this paper, we use solutions of the Killing equation to establish the symmetry of a class of Finslerian spacetime. In particular, we show that the isometric group of a special kind of $(\alpha,\beta)$ space is equivalent to the symmetry of the VSR.

\section{Killing vector in Riemann space}
In this section, we give a brief review of the Killing vectors in Riemann space (further material can be found, for example, in \cite{Weinberg}). Under a given coordinate transformation $x\rightarrow\bar{x}$, the Riemannian metric $g_{\mu\nu}(x)$ transforms as
\begin{equation}
\bar{g}_{\mu\nu}(\bar{x})=\frac{\partial x^\rho}{\partial \bar{x}^\mu}\frac{\partial x^\sigma}{\partial \bar{x}^\nu}g_{\rho\sigma}(x).
\end{equation}
Any transformation $x\rightarrow\bar{x}$ is called isometry if and only if the transformation of the metric $g_{\mu\nu}(x)$ satisfies
\begin{equation}
\label{isometry}
g_{\mu\nu}(\bar{x})=\frac{\partial x^\rho}{\partial \bar{x}^\mu}\frac{\partial x^\sigma}{\partial \bar{x}^\nu}g_{\rho\sigma}(x).
\end{equation}
One can check that the isometric transformations do form a group. It is convenient to investigate the isometric transformation under the infinitesimal coordinate transformation
\begin{equation}
\label{coordinate tran}
\bar{x}^\mu=x^\mu+\epsilon V^\mu,
\end{equation}
where $|\epsilon|\ll1$.
To first order in $|\epsilon|$, the equation (\ref{isometry}) reads
\begin{equation}
V^\kappa\frac{\partial g_{\mu\nu}}{\partial x^\kappa}+g_{\kappa\mu}\frac{\partial V^\kappa}{\partial x^\nu}+g_{\kappa\nu}\frac{\partial V^\kappa}{\partial x^\mu}=0.
\end{equation}
By making use of the covariant derivatives with respect to Riemannian connection, we can write the above equation as
\begin{equation}
\label{killing}
V_{\mu|\nu}+V_{\nu|\mu}=0,
\end{equation}
where $``|"$ denotes the covariant derivative. Any vector field $V_{\mu}$ satisfies equation (\ref{killing}) is called Killing vector.
Thus, the problem of finding all isometries of a given metric $g_{\mu\nu}(x)$ is reduced to find the dimension of the linear space formed by Killing vectors.

In Riemann geometry, by making use of the covariant derivative, one could obtain the Ricci identities or interchange formula
\begin{equation}
\label{Ricci}
V_{\rho|\mu|\nu}-V_{\rho|\nu|\mu}=-V_\sigma R^{~\sigma}_{\rho~\nu\mu},
\end{equation}
where $R^{~\sigma}_{\rho~\nu\mu}$ is the Riemannian curvature tensor.
And the first Bianchi identity for the Riemannian curvature tensor gives
\begin{equation}
\label{cyclic sum}
R^{~\sigma}_{\rho~\nu\mu}+R^{~\sigma}_{\nu~\mu\rho}+R^{~\sigma}_{\mu~\rho\nu}=0.
\end{equation}
Deducing from the equation (\ref{Ricci}) and (\ref{cyclic sum}), we obtain
\begin{equation}
V_{\rho|\mu|\nu}=V_\sigma R^{~\sigma}_{\nu~\mu\rho}.
\end{equation}
Thus, all the derivatives of $V_\mu$ will be determined by the linear combinations of $V_\mu$ and $V_{\mu|\nu}$. Once the $V_\mu$ and $V_{\mu|\nu}$ at an arbitrary point of Riemannian space is given, then $V_\mu$ and $V_{\mu|\nu}$ at any other point is determined by integration of the system of ordinary differential equations. Therefore, the dimension of linear space formed by Killing vector can be at most $\frac{n(n+1)}{2}$ in $n$ dimensional Riemannian space.
If a metric admits that the maximum number $\frac{n(n+1)}{2}$ of Killing vectors, its Riemann space must homogeneous and isotropic (or the space is isotropic for every point). Such space is called maximally symmetry space. In Riemann geometry, the Schur's lemma tells us that a Riemannian space with at least 3 dimension is maximally symmetry space if and only if its sectional curvature is constant. Also, one can check that a 2 dimensional Riemannian space is maximally symmetry space if and only if its sectional curvature is constant. Thus, the maximal symmetry of a given metric is an intrinsic property, and not depending on the choice of coordinate system.

One special maximally symmetry space is the Minkowskian space. The Killing equation (\ref{killing}) of a given Minkowskian metric $\eta_{\mu\nu}(x)$ reduces to
\begin{equation}
\frac{\partial V_\mu}{\partial x^\nu}+\frac{\partial V_\nu}{\partial x^\mu}=0.
\end{equation}
The solution of the above equation is
\begin{equation}
\label{killing s}
V^{\mu}=Q^\mu_{~\nu} x^\nu+C^\mu,
\end{equation}
where $Q_{\mu\nu}=\eta_{\rho\mu}Q^\rho_{~\nu}$ is an arbitrary constant skew-symmetric matrix and $C^\mu$ is an arbitrary constant vector. Thus, substituting the solution (\ref{killing s}) into the coordinate transformation (\ref{coordinate tran}) we obtain
\begin{equation}
\bar{x}^\mu=(\delta^\mu_\nu+\epsilon Q^\mu_{~\nu})x^\nu+\epsilon C^\mu.
\end{equation}
One should find that the term $\delta^\mu_\nu+\epsilon Q^\mu_{~\nu}$ in above equation is just the Lorentz transformation matrix and the term $\epsilon C^\mu$ is related to the spacetime translation. Expanding the matrix $\delta^\mu_\nu+\epsilon Q^\mu_{~\nu}$ and the vector $\epsilon C^\mu$ near identity, one could obtain the famous Poincare algebra.

Other two types of maximally symmetry space are spherical and hyperbolic case. Without loss of generality, we set its constant sectional curvature to be $\pm1$ for spherical and hyperbolic case respectively. The length element of spherical and hyperbolic case is given in a unified form
\begin{equation}
ds^2=\frac{\sqrt{(1+k(x\cdot x))(dx\cdot dx)-k(x\cdot dx)^2}}{1+k(x\cdot x)},
\end{equation}
where the $\cdot$ denotes the inner product with respect to Minkowskian metric and $k=\pm1$ for spherical and hyperbolic case respectively. The metric is given as
\begin{equation}
g_{\mu\nu}=\left(\frac{\eta_{\mu\nu}}{1+k(x\cdot x)}-k\frac{x_\mu x_\nu}{(1+k(x\cdot x))^2}\right),
\end{equation}
where $x_\mu\equiv\eta_{\mu\nu}x^\nu$.
The Christoffel symbols of the above length element is given as
\begin{equation}
\gamma^\rho_{\mu\nu}=-k\frac{x_\mu\delta^\rho_\nu+x_\nu\delta^\rho_\mu}{1+k(x\cdot x)}.
\end{equation}
Thus, the Killing equation (\ref{killing}) now reads
\begin{equation}
\frac{\partial V_\mu}{\partial x^\nu}+\frac{\partial V_\nu}{\partial x^\mu}+\frac{2k}{1+k(x\cdot x)}(x_\mu V_\nu+X_\nu V_\mu)=0.
\end{equation}
The solution of the above equation is
\begin{equation}
\label{k in curv}
V^\mu\equiv g^{\mu\nu}V_\nu=Q^\mu_{~\nu}x^\nu+C^\mu+k(x\cdot C)x^\mu,
\end{equation}
where the index of $Q$ and $C$ are raise and lower by Minkowskian metric $\eta^{\mu\nu}$ and its matrix reverse $\eta_{\mu\nu}$.

\section{Killing vectors in Finsler space}
Instead of defining an inner product structure over the tangent bundle in Riemann geometry, Finsler geometry is base on
the so called Finsler structure $F$ with the property
$F(x,\lambda y)=\lambda F(x,y)$ for all $\lambda>0$, where $x$ represents position
and $y\equiv\frac{dx}{d\tau}$ represents velocity. The Finsler metric is given as\cite{Book
by Bao}
 \begin{equation}
 g_{\mu\nu}\equiv\frac{\partial}{\partial
y^\mu}\frac{\partial}{\partial y^\nu}\left(\frac{1}{2}F^2\right).
\end{equation}
Finsler geometry has its genesis in integrals of the form
\begin{equation}
\label{integral length}
\int^r_sF(x^1,\cdots,x^n;\frac{dx^1}{d\tau},\cdots,\frac{dx^n}{d\tau})d\tau~.
\end{equation}
So the Finsler structure represents the length element of Finsler space.

Like Riemannian case, to investigate the Killing vector we should construct the isometric transformation of Finsler structure.
Let us consider the coordinate transformation (\ref{coordinate tran}) together with the corresponding transformation for $y$
\begin{equation}
\label{coordinate tran1}
\bar{y}^\mu=y^\mu+\epsilon\frac{\partial V^\mu}{\partial x^\nu}y^\nu.
\end{equation}
Under the coordinate transformation (\ref{coordinate tran}) and (\ref{coordinate tran1}), to first order in $|\epsilon|$, we obtain the expansion of the Finsler structure,
\begin{equation}
\label{coordinate tran F}
\bar{F}(\bar{x},\bar{y})=\bar{F}(x,y)+\epsilon V^\mu\frac{\partial F}{\partial x^\mu}+\epsilon y^\nu\frac{\partial V^\mu}{\partial x^\nu}\frac{\partial F}{\partial y^\mu},
\end{equation}
where $\bar{F}(\bar{x},\bar{y})$ should equal $F(x,y)$.
Under the transformation (\ref{coordinate tran}) and (\ref{coordinate tran1}), a Finsler structure is called isometry if and only if
\begin{equation}
F(x,y)=\bar{F}(x,y).
\end{equation}
Then, deducing from the (\ref{coordinate tran F}) we obtain the Killing equation $K_V(F)$ in Finsler space
\begin{equation}
\label{killing F}
K_V(F)\equiv V^\mu\frac{\partial F}{\partial x^\mu}+y^\nu\frac{\partial V^\mu}{\partial x^\nu}\frac{\partial F}{\partial y^\mu}=0.
\end{equation}

Searching the Killing vectors for general Finsler structure is difficult. Here, we give the Killing vectors for a class of Finsler space-$(\alpha,\beta)$ space\cite{Shen} with metric defining as
\begin{eqnarray}
F=\alpha\phi(s),~~~s=\frac{\beta}{\alpha},\\
\alpha=\sqrt{g_{\mu\nu}y^\mu y^\nu}~~{\rm and}~~ \beta=b_\mu(x)y^\mu,
\end{eqnarray}
where $\phi(s)$ is a smooth function, $\alpha$ is a Riemannian metric and $\beta$ is one form.
Then, the Killing equation (\ref{killing F}) in $(\alpha,\beta)$ space reads
\begin{eqnarray}
0&=&K_V(\alpha)\phi(s)+\alpha K_V(\phi(s))\nonumber\\\label{killing ori}
 &=&\left(\phi(s)-s\frac{\partial \phi(s)}{\partial s}\right)K_V(\alpha)+\frac{\partial\phi(s)}{\partial s}K_V(\beta).
\end{eqnarray}
And by making use of the Killing equation (\ref{killing F}), we obtain
\begin{eqnarray}
K_V(\alpha)&=&\frac{1}{2\alpha}(V_{\mu|\nu}+V_{\nu|\mu})y^\mu y^\nu,\\
K_V(\beta)&=&\left(V^\mu\frac{\partial b_\nu}{\partial x^\mu}+b_\mu\frac{\partial V^\mu}{\partial x^\nu}\right)y^\nu,
\end{eqnarray}
where $``|"$ denotes the covariant derivative with respect to the Riemannian metric $\alpha$.
The solutions of the Killing equation (\ref{killing F}) have three viable scenarios. The first one is
\begin{equation}
\phi(s)-s\frac{\partial \phi(s)}{\partial s}=0~~{\rm and}~~ K_V(\beta)=0,
\end{equation}
which implies $F=\lambda\beta$ for all $\lambda\in\mathbb{R}$. The second one is
\begin{equation}
\frac{\partial\phi(s)}{\partial s}=0~~{\rm and}~~K_V(\alpha)=0,
\end{equation}
which implies $F=\lambda\alpha$ for all $\lambda\in\mathbb{R}$. The above two scenarios are just trivial space, we will not consider in this section. Here we mainly consider the case that $\phi(s)-s\frac{\partial \phi(s)}{\partial s}\neq0$ and $\frac{\partial\phi(s)}{\partial s}\neq0$. This will induce the last scenario.
 Apparently, in the last scenario we have such solutions
\begin{eqnarray}
\label{killing F1}
V_{\mu|\nu}+V_{\nu|\mu}&=&0,\\
\label{killing F2}
V^\mu\frac{\partial b_\nu}{\partial x^\mu}+b_\mu\frac{\partial V^\mu}{\partial x^\nu}&=&0,
\end{eqnarray}
while $a_0\neq0$. The first equation (\ref{killing F1}) is no other than the Riemannian Killing equation (\ref{killing}). The second equation (\ref{killing F2}) can be regarded as the constraint for the Killing vectors that satisfy the Killing equation (\ref{killing F1}). Therefore, in general the dimension of the linear space formed by Killing vectors of $(\alpha,\beta)$ space is lower than the Riemannian one.
Additional solutions of (\ref{killing F}) should be obtained from (\ref{killing F}), it will be discussed in next section.

\section{Symmetry of VSR}
One important physical example of $(\alpha,\beta)$ space is VSR. While we take $\phi(s)=s^m$ and $m$ is an arbitrary constant, the Finsler structure takes the form proposed by Gibbons {\it et al}.\cite{Gibbons}
\begin{eqnarray}
\label{vsr metric}
F&=&\alpha^{1-m}\beta^m\nonumber\\
 &=&(\eta_{\mu\nu}y^\mu y^\nu)^{(1-m)/2}(b_\rho y^\rho)^m.
\end{eqnarray}
We denote it by VSR metric.
In this example, $\eta_{\mu\nu}$ is Minkowskian metric and $b_\rho$ is a constant vector. One immediately obtain from the first Killing equation (\ref{killing F1}) of $(\alpha,\beta)$ space,
\begin{equation}
V^{\mu}=Q^\mu_{~\nu} x^\nu+C^\mu.
\end{equation}
And the second Killing equation (\ref{killing F2}) gives the constraint for Killing vector $V^\mu$,
\begin{equation}
b_\mu Q^\mu_{~\nu}=0.
\end{equation}
Taking the light cone coordinate \cite{Kogut} $\alpha=\sqrt{2y^+ y^--y^i y^i}$ (with i ranging over the values 1 and 2) and supposing $b_\mu=\{0,0,0,b_-\}$, we know that in general $Q^{-}_{~\mu}\neq0$. It means the Killing vectors of VSR metric (\ref{vsr metric}) do not have components $Q_{+-}$ and $Q_{+i}$. In another word, the generators of the group that leaves the metric (\ref{vsr metric}) invariant are just four spacetime translation generators $(P_+,P_i,P_-)$ and three subgenerators of Lorentz algebra $(M_{-i},M_{ij})$. This result implies that the VSR metric is invariant under the E(2) group (the group of two-dimensional Euclidean motion).

The above investigation and the Killing equations (\ref{killing F1}) and (\ref{killing F2}) obtained in section 3 are under the premise that the direction of $y^\mu$ is arbitrary. It means that no preferred direction exists in spacetime. If the spacetime does have a special direction, the Killing equation (\ref{killing ori}) will have a special solution. The VSR metric is first suggested by Bogoslovsky \cite{Bogoslovsky}. He assumed that the spacetime has a preferred direction. Following the assumption and taking the null direction to be the preferred direction, we deduce from Killing equation (\ref{killing ori}) that
\begin{eqnarray}
0&=&s^m\left(\frac{1-n}{2\alpha}\left(\frac{\partial V_\mu}{\partial x^\nu}+\frac{\partial V_\nu}{\partial x^\mu}\right)y^\mu y^\nu+ms^{-1}b_\rho\frac{\partial V^\rho}{\partial x^\kappa} y^\kappa\right)\nonumber\\
 &=&s^m\frac{1}{\alpha\beta}\Bigg(\frac{1-n}{2}\left(\frac{\partial V_\mu}{\partial x^\nu}+\frac{\partial V_\nu}{\partial x^\mu}\right)b_\kappa\nonumber\\
 &&+m\eta_{\mu\nu}b^\rho\frac{\partial V_\rho}{\partial x^\kappa}\Bigg)y^\mu y^\nu y^\kappa.
\end{eqnarray}
The above equations has a special solution
\begin{equation}
V_+=(Q_{+-}+m\eta_{+-})x^-+C_+,
\end{equation}
where $Q_{+-}$ is not only an antisymmetrical matrix, but also satisfying the property
\begin{equation}
b_-=-b^+ Q_{+-}.
\end{equation}
It implies that the Lorentz transformation for $b^+$ is
\begin{equation}
\left(\delta^+_++\epsilon(m\delta^+_++Q^+_{~+})\right)b^+=\left(1+\epsilon(n+1)\right)b^+,
\end{equation}
which means the null direction $b^+$ (or $b_-$) is invariant under the Lorentz transformation.
Therefore, if the spacetime has a preferred direction in null direction, the symmetry corresponded to $Q_{+-}$ is restored. In such case the VSR metric is invariant under the the transformations of the group DISIM$_b(2)$ proposed by Gibbons {\it et al}.\cite{Gibbons}.

\vspace{1cm}
Another important physical example of $(\alpha,\beta)$ space is Randers space\cite{Randers}. While we set $\phi(s)=1+s$, the Finsler structure takes the form
\begin{equation}
F=\alpha+\beta.
\end{equation}
Then, in Randers space the Killing equation (\ref{killing ori}) reads
\begin{equation}
\label{Randers killing}
K_V(\alpha)+K_V(\beta)=0.
\end{equation}
Since the $K_V(\alpha)$ contains irrational term of $y^\mu$ and $K_V(\beta)$ only contains rational term of $y^\mu$, therefore the equation (\ref{Randers killing}) satisfies if and only if $K_V(\alpha)=0$ and $K_V(\beta)=0$.
If Randers space is flat, its Killing vectors satisfy the same Killing equations with VSR metric. It implies the flat Randers metric is invariant under the E(2) group semi-direct product with the spacetime translation. Thus, let us focus on the non-flat Randers space. In section 2, we know that a Riemannian space is maximally symmetry space if and only if it is a constant sectional curvature space. Due to the first Killing equation (\ref{killing F1}) of Finslerian space is the same with the Riemannian one and the Killing vectors of Finslerian space must satisfy the constraint (\ref{killing F2}), we know that a Finsler space with constant flag curvature (the notion of flag curvature is the counterpart of sectional curvature) has less independent Killing vectors than its Riemannian counterpart, in general.

Here, we just investigate a special Randers space with constant flag curvature $-\frac{1}{4}$. It is called Funk metric\cite{Funk}. This metric space can be regarded as a flat Minkowskian space which influenced by a radial ``wind" (vector)\cite{Gibbons1}. For Randers space with other types of constant flag curvature, its $\alpha$ term is not a Riemannian constant sectional curvature space. Then, its independent Killing vectors should less than Funk one in general. The Funk metric is given by
\begin{equation}
\label{Funk metric}
F=\frac{\sqrt{(1-x\cdot x)(y\cdot y)+(x\cdot y)^2}}{1-x\cdot x}+\frac{x\cdot y}{1-x\cdot x},
\end{equation}
where the $\cdot$ denotes the inner product with respect to Minkowskian metric. By making use of the result (\ref{k in curv}) obtained in section 2, we know that the first Killing equation $K_V(\alpha)=0$ (\ref{killing F1}) implies
\begin{equation}
\label{k in curv1}
V^\mu=Q^\mu_{~\nu}x^\nu+C^\mu-(x\cdot C)x^\mu.
\end{equation}
And the Funk metric implies $b_\mu(x)=\frac{x_\mu}{1-x\cdot x}$. Then, we obtain the partial derivative for $b_\mu(x)$,
\begin{equation}\label{partial b}
\frac{\partial b_\nu}{\partial x^\mu}=\frac{\eta_{\mu\nu}}{1-x\cdot x}+\frac{2x_\mu x_\nu}{(1-x\cdot x)^2},
\end{equation}
and the partial derivative for Killing vectors $V^\mu$,
\begin{equation}\label{partial V}
\frac{\partial V^\mu}{\partial x^\nu}=Q^\mu_{~\nu}-\delta^\mu_{\nu}(x\cdot C)-x^\mu C_\nu.
\end{equation}
By making use of the equation (\ref{k in curv1}), (\ref{partial b}) and (\ref{partial V}), and deducing from the second Killing equation (\ref{killing F2}) we obtain
\begin{equation}
\label{funk q and c}
\frac{2Q_{\nu\mu}x^\mu}{1-x\cdot x}+C_\nu=0.
\end{equation}
Substituting the equation (\ref{funk q and c}) into (\ref{k in curv1}), we get
\begin{equation}
V^\mu=Q^\mu_{~\nu}x^\nu\left(\frac{3-x\cdot x}{1-x\cdot x}\right).
\end{equation}
Therefore, the dimension of the linear space formed by the Killing vectors of Funk metric is $6$. And the spacetime translation generators corresponded to $C^\mu$ depend on the generators of Lorentz group corresponded to $Q^\mu_{~\nu}$.

\section{Conclusion}
In the past few years, two interesting theories which investigated the violation of  Lorentz Invariance (LI) are proposed. We showed that the Killing vectors satisfy the same Killing equation (\ref{killing F1}) of a Riemannian metric, and the major difference is the Killing vectors of $(\alpha,\beta)$ metric need to satisfy the constraint (\ref{killing F2}). We proved that the isometric group of a flat $(\alpha,\beta)$ space is just the symmetry in VSR proposed by Cohen and Glashow. If the space do not have a preferred direction, we showed that the generators induced by such Killing vectors are just isomorphic to the E(2) group semidirect the spacetime translation. While a preferred direction has chosen, it is showed that the symmetry of such space isomorphic to the SIM(2) group semidirect the spacetime translation. For non flat case, we considered the Funk metric with constant flag curvature. It is showed that the numbers of independent Killing vectors of Funk metric is just $6$. The determination for the maximal number of independent Killing vectors of $(\alpha,\beta)$ space (or general Finsler space) still is a open problem. We hope it could be solved in the future.

\vspace{1cm}
\begin{acknowledgments}
We would like to thank Prof. C. J. Zhu for useful discussions. The
work was supported by the NSF of China under Grant No. 10525522,
10875129 and 10771004.
\end{acknowledgments}


\begin{thebibliography}{999}
\bibitem{Amelino1}G. Amelino-Camelia, Phys. Lett. B {\bf 510}, 255
                  (2001).
\bibitem{Amelino2}G. Amelino-Camelia, Int. J. Mod. Phys. D {\bf 11}, 35
                  (2002).
\bibitem{Amelino3}G. Amelino-Camelia, Nature {\bf 418}, 34
                  (2002).
\bibitem{Smolin1}J. Magueijo and L. Smolin, Phys. Rev. Lett. {\bf
                 88}, 190403 (2002).
\bibitem{Smolin2}J. Magueijo and L. Smolin, Phys. Rev. D {\bf
                 67}, 044017 (2003).
\bibitem{Glashow}A. G. Cohen and S.L. Glashow, Phys. Rev. Lett. {\bf 97}, 021601 (2006).
\bibitem{Kogut}J. B. Kogut and D. E. Soper,  Phys. Rev. D {\bf 1}, 2901 (1970).
\bibitem{Girelli}F. Girelli, S. Liberati and L. Sindoni, Phys. Rev. D {\bf 75}, 064015 (2007).
\bibitem{Gibbons}G. W. Gibbons, J. Gomis and C. N. Pope, Phys. Rev. D {\bf 76}, 081701
(2007).
\bibitem{Finsler DM}Z. Chang and X. Li, Phys. Lett. B {\bf 668}, 453 (2008).
\bibitem{Anderson}J. D. Anderson {\it et al}., Phys. Rev.
Lett. {\bf 81} 2858, (1998); Phys. Rev. D {\bf 65} 082004, (2002);
Mod. Phys. Lett. A {\bf 17} 875, (2002).
\bibitem{Finsler PA}X. Li and Z. Chang, arXiv:gr-qc/0909.3713.
\bibitem{Krasinsky}G. A. Krasinsky and V. A. Brumberg, Celest. Mech. Dyn. Astrn. {\bf 90}, 267 (2004).
\bibitem{Standish}E. M. Standish, Proc. IAU Colloq. {\bf 196}, 163 (2005).
\bibitem{Williams}J. G. Williams and D. H. Boggs, in Proceedings of 16th International Workshop on Laser
Ranging ed. S. Schillak, (Space Research Centre, Polish Academy of Sciences), 2009.
\bibitem{Finsler AU}X. Li and Z. Chang, arXiv:gr-qc/0911.1890.
\bibitem{Killing}W. Killing, J. f. d. reine u. angew. Math. (Crelle), {\bf 109}, 121 (1892).
\bibitem{Weinberg}S. Weinberg, {\it Gravitation and Cosmology: Principles and Applications of
the General Theory of Relativity}, Wiley, New York, 1972.
\bibitem{Book by Bao}D. Bao, S. S. Chern and Z. Shen, {\it An
Introduction to Riemann--Finsler Geometry}, Graduate Texts in
Mathmatics {\bf 200}, Springer, New York, 2000.
\bibitem{Shen}Z. Shen, {\it Some perspectives in Finsler geometry}, MSRI Publication Series. Cambridge: Cambridge university press, 2004.
\bibitem{Bogoslovsky}G. Bogoslovsky, arXiv:gr-qc/0706.2621.
\bibitem{Randers}G. Randers, Phys. Rev. {\bf 59}, 195 (1941).
\bibitem{Funk}P. Funk, Math. Ann. {\bf 101}, 226 (1929).
\bibitem{Gibbons1}G. W. Gibbons, {\it et al.}, Phys. Rev. D {\bf 79}, 044022 (2009).












\end{thebibliography}
\end{document}